\documentclass[a4paper,11pt]{article} \pdfoutput=1 

\usepackage{jheppub} 
\usepackage[T1]{fontenc} 
\usepackage{multirow} 
\usepackage{graphicx}  
\usepackage{mathtools} 

\usepackage[capitalise,english]{cleveref}
\usepackage{xcolor} \usepackage{amsmath} \usepackage{tensor}
\usepackage[utf8]{inputenc} \usepackage{caption} \usepackage{subcaption}

\newcommand{\rar}{\rightarrow} \newcommand{\beq}{\begin{equation}} \newcommand{\eeq}{\end{equation}}
\newcommand{\bc}{\begin{center}} \newcommand{\ec}{\end{center}}

\newcommand{\bea}{\begin{eqnarray}} \newcommand{\eea}{\end{eqnarray}}

\newcommand{\A}{\mathcal{A}} 
\newcommand{\M}{\mathcal{M}} 
\newcommand{\LL}{\mathcal{L}}
\newcommand{\OO}{\mathcal{O}}
\newcommand{\WW}{\mathcal{W}}
\newcommand{\VV}{\mathcal{V}}
\newcommand{\thetah}{\hat{\theta}}
\newcommand{\SM}{\text{SM}}
\newcommand{\BSM}{\text{BSM}}

\newcommand{\lp}{\left(} \newcommand{\rp}{\right)} \renewcommand{\[}{\left[}
\renewcommand{\]}{\right]}

\title{Jet Substructure Measurements of Interference in Non-Interfering SMEFT Effects}
\preprint{MITP/19-011} \author[a]{Rafael Aoude}\author[a,b]{and William Shepherd}

\emailAdd{aoude@uni-mainz.de} \emailAdd{shepherd@shsu.edu}
\affiliation[a]{PRISMA Cluster of Excellence \& Mainz Institute of Theoretical Physics, \\ Johannes
Gutenberg-Universit\"at Mainz, 55099 Mainz, Germany}
\affiliation[b]{Physics Department, Sam Houston State University, Huntsville, TX 77341, USA}

\abstract{
The tails of diboson production at the LHC are sensitive to the interference between Standard Model and higher dimension operators parameterizing the effects of heavy new physics.
However, helicity selection rules for the diboson scattering amplitudes set an obstruction to the na\"ive interference contributions of dimension six
operators, causing the total diboson rate correction's leading contribution to cancel.
In this case, carefully measuring the azimuthal decay angles ``resurrects'' the interference, recouping sensitivity to the ``non-interfering'' operators. We explore these signatures in detail, and find that the EFT uncertainties associated with higher-dimensional operators are uniquely well-suppressed by the construction of an asymmetry variable which is only generated by these non-interfering operators, relegating the effects of higher-dimensional, interfering operators to the same status as statistical errors in this observable.
We perform a complete analysis of this azimuthal interference pattern in hadronic decays of W bosons using jet substructure techniques
to tag the bosons and measure their azimuthal decay angles. This technique provides a valuable cross-check to purely-leptonic measurements of interference resurrection in diboson production.
 }
\begin{document}

\maketitle

\section{Introduction}

The LHC physics program has already had a great success in the discovery of the Higgs boson \cite{Chatrchyan:2012xdj,Aad:2012tfa},
completing the Standard Model (SM) and providing the first direct probe of the mechanism of
electroweak symmetry breaking. Unfortunately, the plethora of new particles expected to explain the
otherwise unnatural parameter values needed to fix the Higgs mass and vacuum expectation value (VEV) to
those required by data have not made themselves known to us. Nonetheless, naturalness remains a compelling argument that new physics
ought to exist near the electroweak scale. This circumstance has led to a resurgence of techniques treating the
SM as an effective field theory (EFT), explicitly allowing the existence of higher-dimensional, non-renormalizable
operators to parameterize the potential effects of new physics too heavy to have been seen yet at the LHC.
There are two physically distinct and perfectly reasonable ways of treating the SM as an EFT; the difference
between them hinges on what assumptions are made about the nature of the experimentally measured scalar.

If one treats the 125 GeV scalar $h$ as potentially already having a new physics origin, at least in part, then the
logical expansion to perform is a simultaneous expansion in $\frac{h}{v}$ and $\frac{D}{\Lambda}$, where $h$
is the scalar discovered at the LHC, $v$ the SM Higgs vacuum expectation value (vev), $D$ is a covariant derivative, and $\Lambda$ is the characteristic scale of
new physics (other than that already assumed to be incorporated in $h$). As $v$ is generally comparable to or
below the scale at which measurements are being made to constrain this EFT, it is necessary to expand in relation
to $\Lambda$ only, with all the vev-suppressed terms resummed and treated as a form factor instead. This
treatment is known as the Higgs EFT or HEFT, and has been investigated in depth; for recent status of constraints in this framework, see~\cite{deBlas:2018tjm}.

If instead one is satisfied by the thus-far agreement of the LHC measurements with the SM predictions for the
Higgs boson, it is possible to insist that the $h$ is in fact the Higgs, embedded in an electroweak doublet with
the would-be Goldstone bosons which become the longitudinal modes of the $W^\pm$ and $Z$ bosons. This
assumption forbids the separate series in $\frac{h}{v}$, leaving a simpler EFT with just one characteristic scale.
This approach has come to be called the SMEFT, and is the one we will adopt for this study. For a recent review of this approach, see~\cite{Brivio:2017vri}.

The primary virtue of any EFT treatment is its independence from the underlying UV physics; it is precisely
our agnosticism regarding the precise nature of the solution to the hierarchy problem of the SM which
motivates us to adopt these methods, rather than directly studying the specific model which we find most pleasing, aesthetically or otherwise. If used properly, these EFT techniques will allow for
bounds to be placed on any model of heavy new physics, importantly including models which have not
yet been invented. In order to retain this virtue, it is important that the analyses performed in SMEFT
not make unwarranted UV assumptions, even implicitly. In particular, we need to allow for multiple
different operators to be active at once, and study all those operators which have an impact on the
observable being measured simultaneously.

Much foundational effort has already been invested in the SMEFT, with the full basis of operators
at leading order (dimension 6)
having been sought by many~\cite{Buchmuller:1985jz,Hagiwara:1993ck,Arzt:1994gp,Gounaris:1998ni,Manohar:2006gz,Manohar:2006ga,Barger:2003rs,Giudice:2007fh,Grinstein:2007iv}, and ultimately found~\cite{Grzadkowski:2010es}. Once a complete and non-redundant
basis was known, it became possible to study the bounds which can be placed on EFT operators in
a UV independent way by performing global fits which allow all the Wilson coefficients which are relevant
to the data being considered to vary simultaneously. These analyses at tree level have now been
performed on the full set of precision electroweak data~\cite{Berthier:2016tkq}, on diboson production data from LEP~\cite{Berthier:2015gja},
on measurements of Higgs properties~\cite{Ellis:2014dva,Gauld:2015lmb,Gauld:2016kuu,Hartmann:2015aia}, and combinations of these~\cite{Butter:2016cvz,Biekotter:2018rhp,Almeida:2018cld}. These tree-level
analyses were able to meaningfully constrain the subset of operators which contribute directly to two to two
scattering on the Z and Higgs poles at tree level in the narrow width approximation.

Given this non-redundant basis, it also became possible to study the behavior of the leading EFT effects 
under SM renormalization~\cite{Jenkins:2013zja,Jenkins:2013wua,Alonso:2013hga},
and that work enabled many analyses of EFT effects at one-loop order, such as corrections to 
Higgs, tops and Z decays~\cite{Hartmann:2016pil,Gauld:2015lmb,Hartmann:2015oia,Dawson:2018pyl,Dawson:2018jlg,Zhang:2013xya,Dawson:2018liq,Dedes:2018seb}, 
QCD and EW radiative corrections~\cite{Vryonidou:2018eyv,Dawson:2018liq,Gauld:2016kuu} 
and some processes such as single top production~\cite{Maltoni:2016yxb}, Higgs production in association with a $t\bar{t}$ pair~\cite{Zhang:2016omx},
and next-to-leading order (NLO) QCD effects in anomalous triple gauge couplings and Drell-Yan processes at the LHC~\cite{Dawson:2018dxp,Baglio:2017bfe,Baglio:2018bkm}. These loop-level calculations introduce dependence (at the order of 10\% corrections to the tree-level EFT effect) on new operators which did not contribute at tree level to these observables. They thus make manifest the
need for additional data, beyond that of single on-shell particle production and decay, in order to successfully
constrain or measure the Wilson coefficients of the SMEFT at accuracy comparable to that available
in the data on the poles.

The series expansion in $\Lambda^{-1}$ in the SMEFT behaves differently away from poles than
on them. The dominant beyond SM (BSM) effects on data at a pole arise from insertions
of the Higgs vev in dimension d operators leading to corrections to couplings already present in the SM. All other effects
are suppressed relative to these contributions by the width of the particle being singly produced.
Off the pole, however, other operators are important, and generically scale in relation to the SM contribution as $\frac{E^2}{\Lambda^2}$,
where $E$ is the characteristic energy of the process. The scaling can arise in two different ways: a shift in the SM couplings 
can violate the energy growth cancellation in would-be SM amplitudes, or an operator can explicitly contain a greater number of dynamical fields or derivatives.
This has motivated many studies of EFT effects in the high-energy tails of distributions, most notably those for 
quark compositeness~\cite{Khachatryan:2014cja,Chatrchyan:2012bf}. This growth
greatly increases the signal effect due to the presence of the EFT correction, but it also hints at the
breakdown of our perturbative expansion, necessitating an appropriate treatment of theoretical
errors due to unknown yet-higher-dimensional operators. A consistent approach to these searches
gives notably weaker, but much more theoretically robust, constraints \cite{Alte:2017pme,Alte:2018xgc}.

In studying the potential effects of the SMEFT, it has been found that certain operators are not able
to interfere with the SM in two-to-two scatterings \cite{Azatov:2016sqh}. In particular, the operators $X^3$,
composed of three insertions of a gauge field strength tensor, do not interfere at leading order. These non-interference phenomena 
are due to amplitude total helicity selection rules forcing the final state gauge bosons to have distinct helicities when generated by SM interactions, while their helicities are required to be the same in order for the $X^3$ operator to couple to them. However, interference effects return
once higher-point amplitudes are considered, either due to the decay of unstable particles or the radiation
of an additional gluon or photon \cite{Azatov:2017kzw,Panico:2017frx,Dixon:1993xd}. 
In both of these cases nontrivial angular correlations occur, and in fact
the case of particle decays has angular structure which causes the contribution to again vanish if
the azimuthal decay angles are integrated over, as they normally are for the purposes of counting ``2 to 2''
events treating such an unstable particle as a final state. Thus, accessing the interference effects in diboson production,
for instance, requires measuring these decay angles. This of course benefits greatly from
knowing the full kinematics of the events, which is generally not possible in
leptonic vector boson decays due to the presence of neutrinos in the final state.

Expanding the differential cross-section in powers of $\Lambda$, the first new physics term appearing, at $\Lambda^{-2}$ order, is due to the interference between
SM and BSM amplitude from a dimension-6 operator insertion, followed by the BSM amplitude squared and the interference between a dimension-8 operator (or pair of dimension-6 operators) in one matrix element and the SM amplitude,
at $\Lambda^{-4}$ . 
\beq
	d\sigma = d\sigma_{\SM} + \frac{1}{\Lambda^2} d\sigma_{\SM \times \BSM_6} + \frac{1}{\Lambda^4} \left( d\sigma_{\BSM_6^2} +  d\sigma_{\SM \times \BSM_8} \right) + ...
\eeq
However, since a complete dimension-8 basis for SMEFT is not known, it is best not to treat the $d\sigma_{\BSM_6^2}$ piece as signal for this analysis.
We follow the treatment of~\cite{Alte:2017pme,Alte:2018xgc}, truncating the expansion at order $\Lambda^{-2}$ for the signal and considering the dimension-6 squared term as an estimation of the theory error associated with the neglect of higher orders in the perturbation series in powers of $\Lambda^{-2}$.

Tagging techniques for jets which are due to decaying heavy particles are well advanced and accepted
in both the experimental and theoretical communities \cite{Abdesselam:2010pt,Altheimer:2012mn,Altheimer:2013yza}, and will give access to fully-reconstructable
kinematics in diboson processes. In this article we explore the application of these techniques to measure
the decay angles and improve the resulting reach of LHC data for these operators, while performing for the
first time an analysis of diboson observables which estimates consistently the theory error due to neglecting effects higher order
in $\Lambda^{-1}$.

We find that these angular-only interference effects have multiple unusual properties compared to the more typical
SMEFT analysis; nontrivial azimuthal decay angle effects can only arise from interference effects between bosons of differing helicity,
and even then only interferences between amplitudes for which it is impossible to choose all helicities
to match are able to give rise to the asymmetry which we shall use to perform this analysis. Therefore,
these analyses are sensitive to very few operators, giving a much cleaner signal model and more straightforward
interpretation than usual in terms of constraints on Wilson coefficients. Additionally, this feature leads to the
extremely unusual fact that the errors due to higher-dimensional operators on these observables also
generically do not contribute to these asymmetries beyond correcting the overall diboson rate and thus altering the statistical fluctuations of the asymmetry variable. This stands in stark contrast to the usual state of consistent SMEFT measurements, where the number of possible causes of a deviation from the SM in any given observable is generally vast, and the theoretical uncertainties arising from higher-order EFT contributions is often significant and occasionally the dominant source of errors.

In the next section, we review the arguments for non-interference between certain SMEFT operators
and SM amplitudes, and then investigate the source and nature of the azimuthal correlations that
arise when the decay of the intermediate weak bosons is consistently applied. We then discuss the
jet substructure techniques which we shall use to differentiate between weak bosons and ordinary
QCD jets, and discuss as well how the azimuthal angles are recoverable from the standard tools
already regularly used in substructure-based searches in \cref{sec:substructure}. In \cref{sec:search},
we describe in detail the search design for the LHC which can exploit these azimuthal correlations,
and discuss our signal and background modeling. We then present the resulting sensitivities in \cref{sec:results}
and conclude in \cref{sec:conc}. We include explicit calculations of the two-to-four amplitudes and an exploration
of the impact of na\"\i vely interfering operators in \cref{appendix:2to4,appendix:ops}.

\section{Non-interference in the SMEFT} 
\label{sec:nonint}

We work in the Warsaw basis~\cite{Grzadkowski:2010es}, which has been constructed by systematically using the SM
equations of motion to reduce the number of derivatives in the retained operators in favor of
instead including more fields. This has multiple advantages, but the most important for the purposes
of this article are that it ensures that corrections to propagators are due only to corrections to
input parameters and retain their SM form, and it avoids couplings at higher-point vertices which
have momentum dependence able to cancel the propagator of one of the particles incident on the vertex.

\subsection{Triple-Gauge-Coupling in the SMEFT}
\label{subsec:TGC}
The anomalous TGC Lagrangian, in the SMEFT,  has been historically written as
\beq
\label{aTGC}
\frac{-\LL^{\text{SMEFT}}_{TGC}}{g_{VWW}} = i\bar{g}_1^V ( \WW^+_{\mu\nu} \WW^{-\mu} - \WW^-_{\mu\nu} \WW^{+\mu}) \VV^\nu + i\bar{\kappa}_V \WW^+_\mu \WW^-_\nu \VV^{\mu\nu}
+ i\frac{\bar{\lambda}_V}{\bar{M}^2_W} \VV^{\mu\nu} \WW^{+\rho}_\nu \WW^-_{\rho\mu},
\eeq
where $\WW_{\mu}$ and $\VV_\mu$ are the canonically normalized gauge fields. The couplings above can be written as~\cite{Berthier:2016tkq}
\begin{gather}
g_{AWW} = \hat{e},\quad g_{ZWW} = \hat{e} \cot{\hat{\theta}}, \quad \bar{g}_1^V = g_1^V + \delta g_1^V \nonumber \\
\bar{\kappa}_V = \kappa_V + \delta \kappa_V, \quad \bar{\lambda}_V = \lambda_V + \delta \lambda_V\,,
\end{gather}
where the hat notation indicates a measured coupling constant value. The coupling shifts of the historical Lagrangian form \cref{aTGC} can be directly calculated in terms of the SMEFT Warsaw basis operators, given here using the $\{\hat{\alpha},\hat{G}_F, \hat{m}_Z\}$ input scheme:
\begin{align}
\delta g_1^A&=0,           &  \delta g_1^Z&=\frac{1}{2\sqrt{2}\hat{G}_F} \left( \frac{s_{\thetah}}{c_{\thetah}} + \frac{c_{\thetah}}{s_{\thetah}}  \right) C_{HWB} + \frac{1}{2}\left( \frac{1}{s^2_{\thetah}} +  \frac{1}{c^2_{\thetah}}\right),        \\
\delta \kappa_A&=\frac{1}{\sqrt{2}\hat{G}_F}\frac{c_{\thetah}}{s_{\thetah}} C_{HWB},         &  \delta\kappa_Z&=\frac{1}{2\sqrt{2}\hat{G}_F} \left( -\frac{s_{\thetah}}{c_{\thetah}} + \frac{c_{\thetah}}{s_{\thetah}}  \right) C_{HWB} + \frac{1}{2}\left( \frac{1}{s^2_{\thetah}} +  \frac{1}{c^2_{\thetah}}\right),  \\
\label{eq:deltaCs}
\delta\lambda_A&=6s_{\thetah} \frac{\hat{m}^2_W}{g_{AWW}} C_W,   &  \delta\lambda_Z&=6c_{\thetah} \frac{\hat{m}^2_W}{g_{ZWW}} C_W,   
\end{align}
where the Wilson coefficients correspond to the operators:
\beq
	\mathcal{Q}_{HWB} = H^\dag \tau^I H W^I_{\mu\nu} B^{\mu\nu} \quad \text{and} \quad \mathcal{Q}_W = \epsilon^{IJK} W^{I\nu}_{\mu} W^{J\rho}_{\nu} W^{K\mu}_{\rho},
\eeq

and the SM couplings are $g_1^V = \kappa_V = 1$ and $\lambda_V = 0$. In the chosen input parameter scheme, gauge-invariance at order $\Lambda^{-2}$ requires that $\delta \kappa_Z = \delta g_1^Z - t^2_{\hat{\theta}} \delta\kappa_A$, $\delta g_1^A = 0$ and $\delta\lambda_A = \delta \lambda_Z$~\cite{PhysRevD.48.2182}.

Generic values of the coupling shifts $\delta g_1^V$ and $\delta \kappa_V$ spoil the SM cancellation of amplitude terms that grow with energy in the charged current diagrams, also known as CC03 diagrams, resulting in a growing effect of these coupling corrections in the high-energy tails of the distribution. This is a somewhat unique feature in the SMEFT, as usually the source of growing-with-energy terms is operators which contain extra dynamical fields or derivatives, but in this case operators which are functionally dimension-4 after Higgs vev insertions still lead to growing effects.

On the other hand, the operator $\mathcal{Q}_W$, related to $\bar{\lambda}_V$, couples only to three-boson combinations which all have identical helicities. Since the SM couplings always lead to the production of two opposite-helicity vector bosons at leading order, this leads to a non-interference between SM and the $\mathcal{Q}_W$-induced BSM four-point amplitudes for diboson production.

\subsection{Interference obstruction for $ 2 \rar 2$ processes}

Within a derivative-reduced basis like the Warsaw basis \cite{Grzadkowski:2010es}, it is possible to identify the helicities of particles
which can be coupled to by SM and SMEFT operators in a two-to-two scattering process by constructing
the amplitude out of on-shell three-point sub-amplitudes. 
If there were a contribution that corrected the propagator in a non-standard way or a vertex with momentum dependence which could cancel the
propagator on an incoming particle this would not be possible in general. Investigating the implications of these techniques,
Ref.~\cite{Azatov:2016sqh} finds that there are some classes of operators which do not interfere
with the SM amplitudes at leading order in two-to-two scattering of definite-helicity particles, and
thus have leading-order in cutoff scale interactions which do not grow with energy in the way
generically expected of higher-dimensional operators, instead scaling as $\frac{m^2}{\Lambda^2}$, where $m$ is the mass of the heaviest particle whose helicity can be flipped to allow interference with the SM amplitude.

\begin{figure}[t]
\centering
\includegraphics[width=0.85\linewidth]{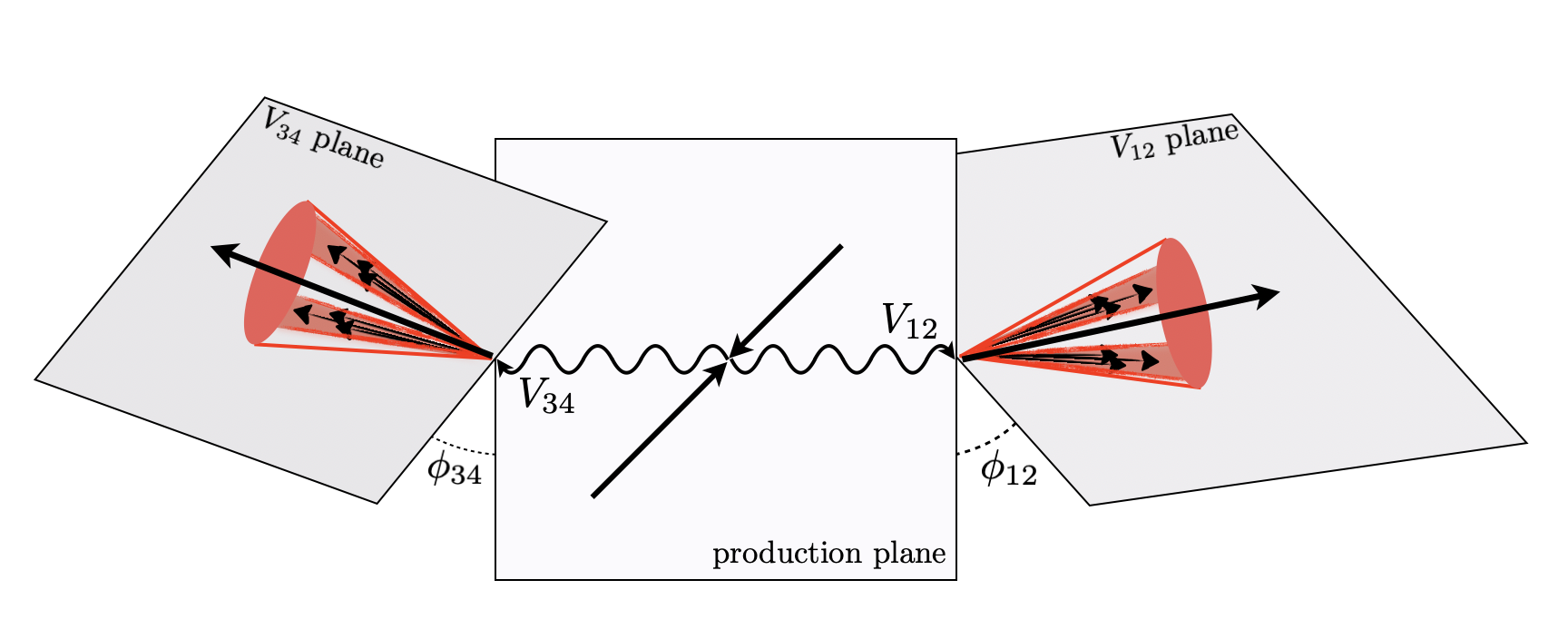} 
\caption{Diboson production planes and boson hadronic decay planes. For the semileptonic channel, one of the fat jets is a dilepton system
but the kinematics is similar. The relevant azimuthal angles for this analysis is the angle between the production plane and the boson decay planes.}
\label{planes}
\end{figure}

The authors recommend to study the effect of the square of the dimension-6 operator amplitude, but the effects at this order in the
inverse cutoff scale expansion are not predicted solely by the dimension-6 basis, and are thus
best treated as theoretical errors rather than signal contributions \cite{Alte:2017pme,Alte:2018xgc}. This non-interference has
long been known in the context of higher-dimensional couplings of gluons \cite{Dixon:1993xd}, with proposals for
how to probe this operator in spite of the non-interference similar to those in the electroweak case.

Multiple investigations of how best to recover from this non-interference effect have been
undertaken since then, focusing on the fact that the weak bosons ultimately are not final-state
particles but also decay \cite{Azatov:2017kzw,Panico:2017frx,Azatov:2019xxn}. Their decays revive the interference effect in two-to-four
scattering in such a way that if the decay kinematics are not explicitly measured the interference
effects again largely cancel.

\subsection{Azimuthal structures in decay angles}

All dependence on azimuthal angles in decays arises from the angular momentum portion of the gauge
boson wavefunction, proportional to $e^{i\lambda\phi}$ for helicity $\lambda$ and decay azimuthal angle $\phi$ relative to the production plane. The azimuthal 
angles and the relevant planes are illustrated in Figure~\ref{planes}, where the final state fermions are taken to be quarks.
When an amplitude of fixed helicity is squared, these phases trivially cancel, but interfering amplitudes with different
helicities will yield trigonometric functions of the decay angles. 
At tree level, the only imaginary contribution to the amplitude comes from the gauge boson decay and any CP violating coupling. Therefore, any CP-conserving operator gives a structure composed of cosines of the azimuthal angle of the decaying bosons. 
On the other hand, a CP-odd operator leads to a combination of sines due to the factor of $i$ that appears on the operator~\cite{VenturiniCPodd}. In this article, we focus on new physics which is not CP-violating, as CP violation is generally strongly constrained by precisely-measured low-energy observables. The inclusion of 
$\mathcal{Q}_{\widetilde{W}}$ generates non-negligible contribution at 1-loop to the neutron electric dipole moment, therefore its Wilson coefficient should be highly
suppressed~\cite{Gripaios:2013lea}.

Whether or not this structure is observable depends on the nature of the event being measured,
though. If we are unable to distinguish experimentally between the two decaying bosons, as is the
case for a fully-hadronic event, then the decay angle is not experimentally defined from $0$ to
$2\pi$ but rather from $0$ to $\pi$, as it is experimentally not possible to orient the production plane in this
case, despite the fact that quantum mechanically it is in principle observable in the full range.
If, additionally, we also cannot tell apart the particle from the antiparticle in the boson decay,
then the angle is actually defined only on the interval from $0$ to $\frac{\pi}{2}$. In order to map
the azimuthal behavior into this observable variable, it is necessary to perform two ``foldings'' of
the distribution in each decay angle, defining the observable angular function $f\left(\phi\right)$ as
\footnote{Here the dependence on the second decay angle is implicit for simplicity.}
\beq 
2f\left(\phi\right)=g\left(\phi\right)+g\left(\pi-\phi\right),\qquad
2g\left(\phi\right)=h\left(\phi\right)+h\left(2\pi-\phi\right), 
\eeq 
where $h\left(\phi\right)$ is
the appropriate trigonometric function resulting from the helicity interference in question.
In this way, $f\left(\phi\right)$ is now defined only in the observable range. 
Some angular distributions $h\left(\phi\right)$ vanish after being folded in this way, 
notably cosines of $(2n+1) \phi$, which are odd under $2\pi$-folding.

We divide the remaining functions into two sets based on their parity under the operation
$\left(\phi_{12},\phi_{34}\right)\to\left(\frac{\pi}{2}-\phi_{12},\frac{\pi}{2}-\phi_{34}\right)$.
The set of functions which is even under this transformations are cosines of angle sums
and arises in interferences of amplitudes which also are able to interfere with identical
helicities. The set of functions odd under this parity are sum of cosines and can arise at
tree level only from the interference of amplitudes which cannot be combined to interfere identical
helicities for both gauge bosons. 

We calculate all the relevant amplitudes for the interference effect we are searching for, the SM
background processes, and the theory uncertainty effects due to the squared BSM amplitude in \cref{appendix:2to4}, and those due to
the effects of other operators in \cref{appendix:ops}. After the proper foldings, the single-insertion SMEFT amplitude is:
\bea 
\frac{d\sigma_{\text{int}}(s)}{d\phi_{12}d\phi_{34}} \sim \lp-\frac{\delta\lambda_Z \,s}{m_W^2}\rp ( \cos 2\phi_{12} + \cos 2\phi_{34}).
\eea

Strikingly, only the effects of na\"\i vely non-interfering operators interfering with the SM give rise to angular structures which are odd under this parity transformation. The purely SM amplitude has azimuthal dependence, but that dependence is even underthis parity, as is the azimuthal dependence of the non-interfering SMEFT amplitude squared or the na\"\i vely-interfering SMEFT operators interfering with the SM. 
As a result, the ability to measure an asymmetry corresponding to these odd distributions probes only the non-interfering operators, which is just $\mathcal{Q}_{\text{W}}$ at leading order in $\sqrt{s}$. No other operator can produce this distribution with comparable growth in the energy tails; they either require a vev-insertion
or occur only with slower energy growth.

\section{Jet substructure with azimuthal decay angles} 
\label{sec:substructure}

In the last decade, particle identification based on substructure techinques has developed explosively, with the appearance of many new techniques~\cite{Abdesselam:2010pt,Altheimer:2012mn,Altheimer:2013yza}. They were mainly developed to distinguish boosted tops and $W/Z/h$ bosons from QCD jets. 
In particular, boosted Higgs analyses led to the development of the BDRS subjet/filter algorithm~\cite{Butterworth:2008iy}, which made feasible the study of 
$pp \rar HV, H \rar b \bar{b}$ at the LHC.
In early studies, the ATLAS and the CMS experiments used the so-called "YSplitter"~\cite{Butterworth:2002tt} and the "Hopkins" top-tagger~\cite{Kaplan:2008ie} to study hadronically top
decay and efficiently distinguish them from QCD jets. Latter, the HEPTopTagger~\cite{Plehn:2009rk} proposed to reconstruct a not so highly boosted top, particularly useful in SUSY searches.

The jet substructure techniques can be broadly divided into two categories: jet grooming and object discrimination. The former one is built in order to remove, as much as possible, initial state radiation, underlying event, and pileup effects from the hard event of interest. The Soft Drop algorithm~\cite{Larkoski:2014wba} is another example of object discrimination, built for use in tagging $W/Z$ hadronic decays at CMS. Since it is known that QCD mostly produces $1$-prong structures while $W/Z$ and tops produces $2$- and $3$-prong jets, it is useful to use a tagging algorithm to discriminate $1$-prong 
from $N$-prong structures, characterizing the discrimination algorithms. In this study, in order to successfully identify the boosted $W$ boson decaying hadronically, we employ the {\it N-subjettiness}~\cite{Thaler:2010tr} algorithm, which
introduces the jet shape variables $\tau_N$:
\beq 
\tau_N = \frac{1}{d_0} \sum_k p_{T,k} \text{min}\{\Delta R_{1,k}, \Delta R_{2,k}, ... , \Delta R_{N,k}\},
\eeq 
where $p_{T,k}$ is the transverse momentum of the particle $k$ and $\Delta R_{J,k} = \sqrt{ (\Delta \eta)^2 + (\Delta \phi)^2 }$ is the rapidity-azimuth 
pseudo-angular distance between the subjet axis candidate $J$ and the jet constituent particle $k$. The normalization is defined as $d_0 = \sum_k p_{T,k} R_0$,
where $R_0$ is the jet radius.

This variable was designed to be minimal whenever the $N$-subjet hypothesis describes the jet well, i.e. there is not a need for more than $N$ subjets, and the normalisation ensures that $\tau_N$ is near one when the jet is poorly described.
However, the ratio $\tau_N/\tau_{N-1}$  actually has a better discrimination power than $\tau_N$ itself. 
Two prong structures are well identified by smaller values of $\tau_{21} \equiv \tau_2/\tau_1$, 
meanwhile the other $\tau$ ratios (e.g $\tau_{32} \equiv \tau_3/\tau_2$) approach one, as both values are comparably small. 
We make use of $\tau_{32}$ to reject top jets and $\tau_{21}$ to
separate decaying electroweak bosons from QCD background events.

In this study, we consider two decay channels: The {\bf semileptonic} channel, where a diboson $W^{\pm}Z$ is produced with the $Z$ decaying 
leptonically and $W^{\pm}$ into a fat jet and the {\bf hadronic} channel, where a $W^+W^-$ or $W^\pm Z$ pair is produced and both intermediate vector bosons decay hadronically. 
In both channels, the final state includes boson-jets arising from the hadronically-decaying $W^\pm$ which are reconstructed initially as `fat jets'. 
In order to look inside them and fully reconstruct the decay kinematics of the vector boson, we utilize N-subjettiness.

\begin{figure}[!htb]
\centering
\includegraphics[width=\linewidth, page = 1]{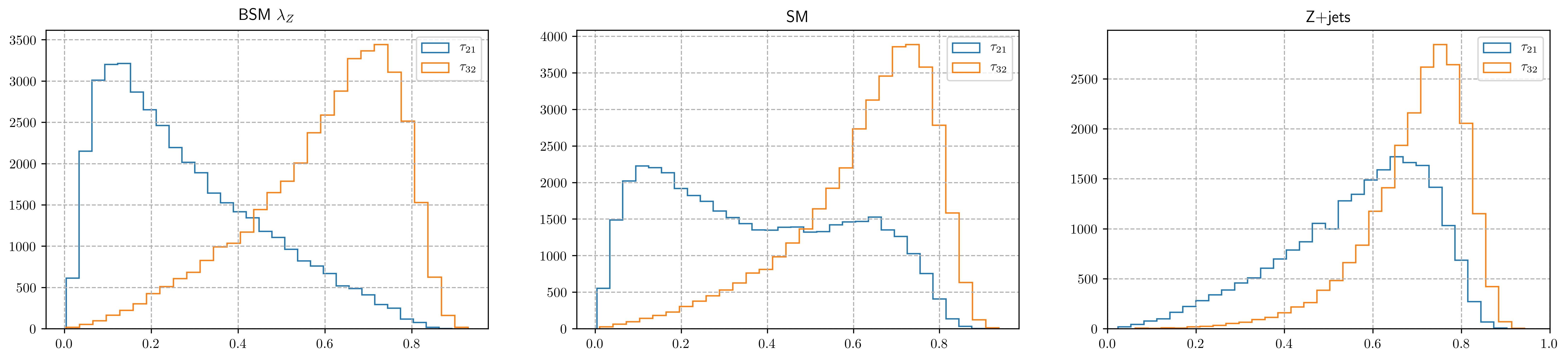} 
\caption{{\it N-subjettiness} ratios $\tau_{21}$ and $\tau_{32}$ for the single fat jet in the Semileptonic channel, plotted separately for the signal case as well as the SM irreducible background diboson processes and the SM QCD background due to $Z$ boson and jet associated production.}
\label{taus_semilep}
\end{figure}

\begin{figure}[!htb]
\centering
\includegraphics[width=\linewidth, page = 1]{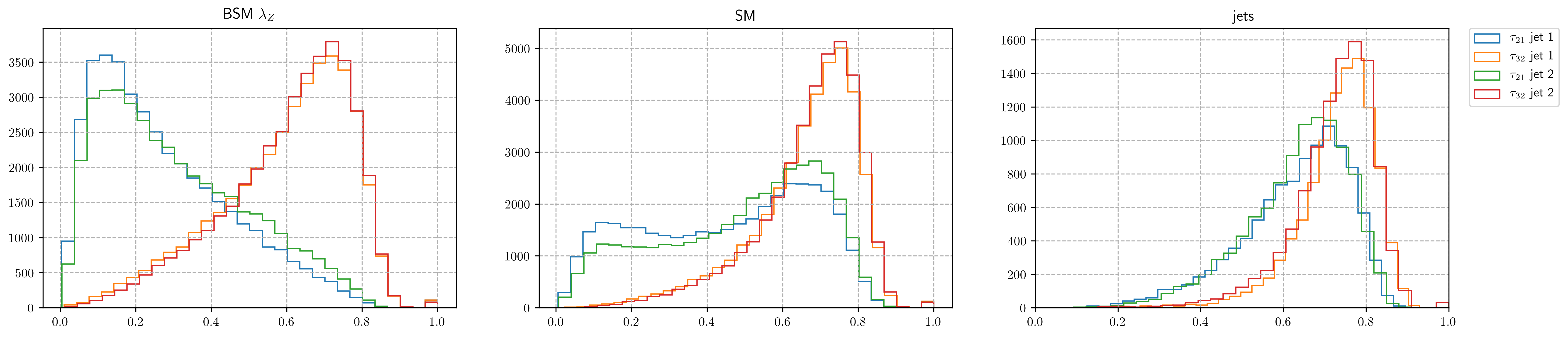} 
\caption{{\it N-subjettiness} ratios $\tau_{21}$ and $\tau_{32}$ for both fat jets in the fully hadronic channel, plotted separately for the signal case as well as the SM irreducible background diboson processes and the SM QCD background due to multijet production.}
\label{taus_had}
\end{figure}

In Figure~\ref{taus_semilep}, we show the ratios $\tau_{21}$ and $\tau_{23}$ of the single $R=1.0$ fat jet produced in the semileptonic channel, for the BSM single insertion 
of $\delta\lambda_Z$, for the diboson SM production and for the SM background $Z$+jets. As we can clearly see, $\tau_{21}$ is small for BSM and SM and peak around $0.6$ for $Z$+jets, which suggests that the two former channels have a $2$-prong structure most of the time, while for the $Z$+jets the $\tau$'s behaviour does not favour this hypothesis. 
In Figure~\labelcref{taus_had}, we see a similar behaviour for the hadronic channel, where we obtain the $2$-prong structure for BSM and SM diboson production but not 
for multijet production. 
The substructure tagging efficiencies for both channels considered in this study are shown in Table~\ref{cuts_had} and~\ref{cuts_semilep}. In both channels we have confirmed that the search design cuts consider in Section~\ref{sec:search} does not affect this behaviour.
Note that the propensity for the BSM signal events to be two-prong-like with low $\tau_{21}$ is even greater than that for the SM diboson events; this arises from
the fact that this interference process effectively causes the polar decay angle to behave in a way reminiscent of longitudinal gauge bosons, 
even though it is the result of interfering transverse bosons of opposite helicity; see \cref{appendix:2to4} for details.

After identifying the tagged events, we need to reconstruct the azimuthal angles of the decay plane. As a proxy for the decay product partons' directions of travel, we utilize the subjet axes which appear in the definition of the $N$-subjettiness. These axes depend on a recombination scheme, for which we utilize One-pass Winner-Take-All (WTA) $k_T$ algorithm; this matches quite well the partonic momenta~\cite{Larkoski:2014uqa}, and we have confirmed explicitly that One-pass E-Scheme $k_T$ Axes do not yield noticeably different results.  

\section{Search design at the LHC} 
\label{sec:search} 
We implemented the relevant operators for this analysis in \texttt{FeynRules}~\cite{Alloul:2013bka} and the simulation for the hadronic and semileptonic channel was
performed using \texttt{MadGraph5}~\cite{Alwall:2014hca}, followed by showering/hadronization via
\texttt{PYTHIA 8}~\cite{Pythia1}.
The fat jets were reconstructed with $R=1.0$ using anti-$k_T$ and E-scheme recombination. 
Each fat jet is then required to have 2 subjets and its axes are reconstructed using WTA $k_T$. 
The subjet reconstruction was done using~\texttt{FastJet}~\cite{Cacciari:2011ma,Cacciari:2005hq}.

The calorimeter cannot resolve tracks that lie within the same cell~$(0.1 \times 0.1)$ in $(\phi \times
\eta)$ and also does not identify soft particles $p_T < 0.5$. We performed the analysis with and without the 
calorimeter granularization and no non-trivial structure is introduced due to the calorimeter; we retain this constraint on our simulated calorimeter to remain 
conservative in our estimates of what sensitivity can be reached.

We also investigated the effects of two techniques of jet grooming for pileup suppression applicable to anti-$k_T$ jets: Jet Trimming ($R_{\text{trim}} = 0.3$ and $p_\text{T,frac} = 0.05$) and Pruning ($R_{\text{cut}} = 0.1$ and $z_{\text{cut}} = 0.05$). The signal events exhibited no relevant modifications in the azimuthal pattern nor in cut efficiencies, indicating reasonable robustness of this analysis technique against pileup effects.

Several backgrounds contribute to diboson production; for the fully-hadronic case, the expected background for
this search consists of multijet events ($pp \rar jj, jjj, jjjj$), SM $W^+W^-$ production, $t\bar{t}$, $W$+ jets and $tW$ production.
In the  semileptonic analysis, we have $Z$+jets, $t\bar{t}$ production and SM $WZ$ production. While none of these backgrounds intrinsically contribute to the asymmetry we search for here, care must be taken to ensure that the analysis cuts do not induce such an asymmetry accidentally; we've explored the effect of each cut and confirmed that they do not cause such azimuthal distortions in the background distributions.

Although the signal diboson cross-section for the hadronic channel is larger than that for the semileptonic due to the greater hadronic branching ratios and inclusion of the $W^+ W^-$ intermediate state, it suffers from more background contamination due to the presence of pure-QCD backgrounds. With a good choice of topology and tagging cuts we can largely eliminate peaking backgrounds without spoiling the 
BSM interference pattern, which are shown in Tables~\ref{cuts_had} and~\ref{cuts_semilep}. Background suppression cuts importantly include an acoplanarity cut to ensure a back-to-back diboson production and a cut requiring a small $p_T$ asymmetry, defined as $(p_{T_1} - p_{T_2})/(p_{T_1} + p_{T_2})$, between the two fat jets (or the fat jet and the dilepton system) in order to remove events with one poorly reconstructed jet. Tagging cuts include a fat jet mass requirement (or dilepton mass in the semileptonic case), small $\tau_2/\tau_1$ for the fat jet, and an $3$-prong structure rejection, which means a large $\tau_3/\tau_2$. The cut values we employ are:
\begin{itemize}
	\item fat jet mass: $40 \leq m_{j} \leq 100$ GeV (hadronic) and $65 \leq m_{j} \leq 105$ GeV (semileptonic)
	\item dilepton mass: $ 80 \leq m_{\ell\ell} \leq 100$ GeV
	\item acoplanarity: $\Delta\phi < 0.5$
	\item p$_\text{T}$ asymmetry: $\Delta p_\text{T}  < 0.15$ 
	\item tagging: $(\tau_2/\tau_1 < 0.45)$ and $(\tau_3/\tau_2 > 0.45)$ for all fat-jets
\end{itemize}

\begin{table}[!ht] 
	\centering 
	\scalebox{0.9}{
	\begin{tabular}{|l|cc|cc|cc|cc|cc|c|} \hline
			 cut                              & \multicolumn{2}{c}{BSM($\delta\lambda_Z$)\,[\%]} \vline& \multicolumn{2}{c}{SM\,[\%]}\vline & \multicolumn{2}{c}{W$+$jets\,[\%]} \vline& \multicolumn{2}{c}{$tW$\,[\%]} \vline & \multicolumn{2}{c}{$t\bar{t}$\,[\%]} \vline& jets\,[\%]\\ \hline \hline 
fat jet mass            & 68.4  & 68.4   & 28.2  & 28.2  & 1.66   & 1.66   & 15.4  & 15.4    & 14.0  & 14.0    & -\\ 
acoplanarity            & 15.3  & 32.9   & 34.9  & 13.1  & 18.76  & 0.77   & 20.3  & 6.1     & 20.6  & 5.3       &-\\ 
p$_\text{T}$ asymmetry  & 89.0  & 31.3   & 76.2  & 11.9  & 35.98  & 0.59   & 38.3  & 4.3     & 39.8  & 3.8       &-\\  
tagging                 & 39.1  & 15.3   & 18.8  & 6.1   & 2.62   & 0.20   & 12.5  & 2.0     & 11.7  & 1.7       &-\\ \hline
total               &\multicolumn{2}{c}{15.3}\vline & \multicolumn{2}{c}{6.1}\vline & \multicolumn{2}{c}{0.20}\vline   & \multicolumn{2}{c}{2.0}\vline & \multicolumn{2}{c}{1.7}\vline & $10^{-3}$ \\\hline
	\end{tabular} }
	\caption{Efficiency table for topology and tagging cuts for the hadronic case for the center-of-mass-energy $0.5$ TeV  $\leq \sqrt{\hat{s}} \leq 2.1$ TeV.
	 The first (second) column of each background represent the individual (sequential) cut efficiency.
	 For the jets background, we assumed the ATLAS~\cite{Aaboud:2017eta} efficiency; note that our tagging efficiency for bosons is actually slightly below the ATLAS number, so this efficiency is a conservative estimate.} 
	\label{cuts_had} 
	\end{table}

\begin{table}[!ht] 
	\centering 
	\scalebox{1.0}{
	\begin{tabular}{|l|cc|cc|cc|cc|} 	\hline
		cut  & \multicolumn{2}{c}{BSM($\delta\lambda_Z$)\,[\%]}\vline & \multicolumn{2}{c}{SM\,[\%]}\vline   & \multicolumn{2}{c}{Z$+$jets\,[\%]}\vline & \multicolumn{2}{c}{$t\bar{t}$\,[\%]}\vline \\ \hline \hline 
fat jet and dilepton mass & 60.7 & 60.7 & 31.8 &  31.8 & 7.10  & 7.10 & 7.73   & 7.73 \\ 
acoplanarity              & 47.8 & 29.9 & 30.8 &  12.1 & 7.10  & 0.57 & 7.99   & 0.62 \\ 
p$_\text{T}$ asymmetry    & 93.5 & 29.2 & 57.1 &  10.8 & 20.46 & 0.08 & 14.13  & 0.04 \\ 
tagging                   & 63.3 & 21.7 & 44.9 &  8.5  & 22.95 & 0.05 & 24.69  & 0.02 \\ \hline
total & \multicolumn{2}{c}{21.7}\vline & \multicolumn{2}{c}{8.5} \vline & \multicolumn{2}{c}{0.05} \vline& \multicolumn{2}{c}{0.02} \vline \\ \hline
	\end{tabular}}
	\caption{Efficiency table for topology and tagging cuts for the semileptonic case for the center-of-mass-energy $0.5$ TeV $\leq \sqrt{\hat{s}} \leq 2.1$ TeV. As in the
	hadronic case, the first (second) column is the individual (sequential) cut efficiency.} 
	\label{cuts_semilep}
\end{table}

\begin{figure}[!htb]
	\centering 
	\begin{subfigure}{.49\textwidth} 
		\centering
		\includegraphics[width=1.1\linewidth, page = 1]{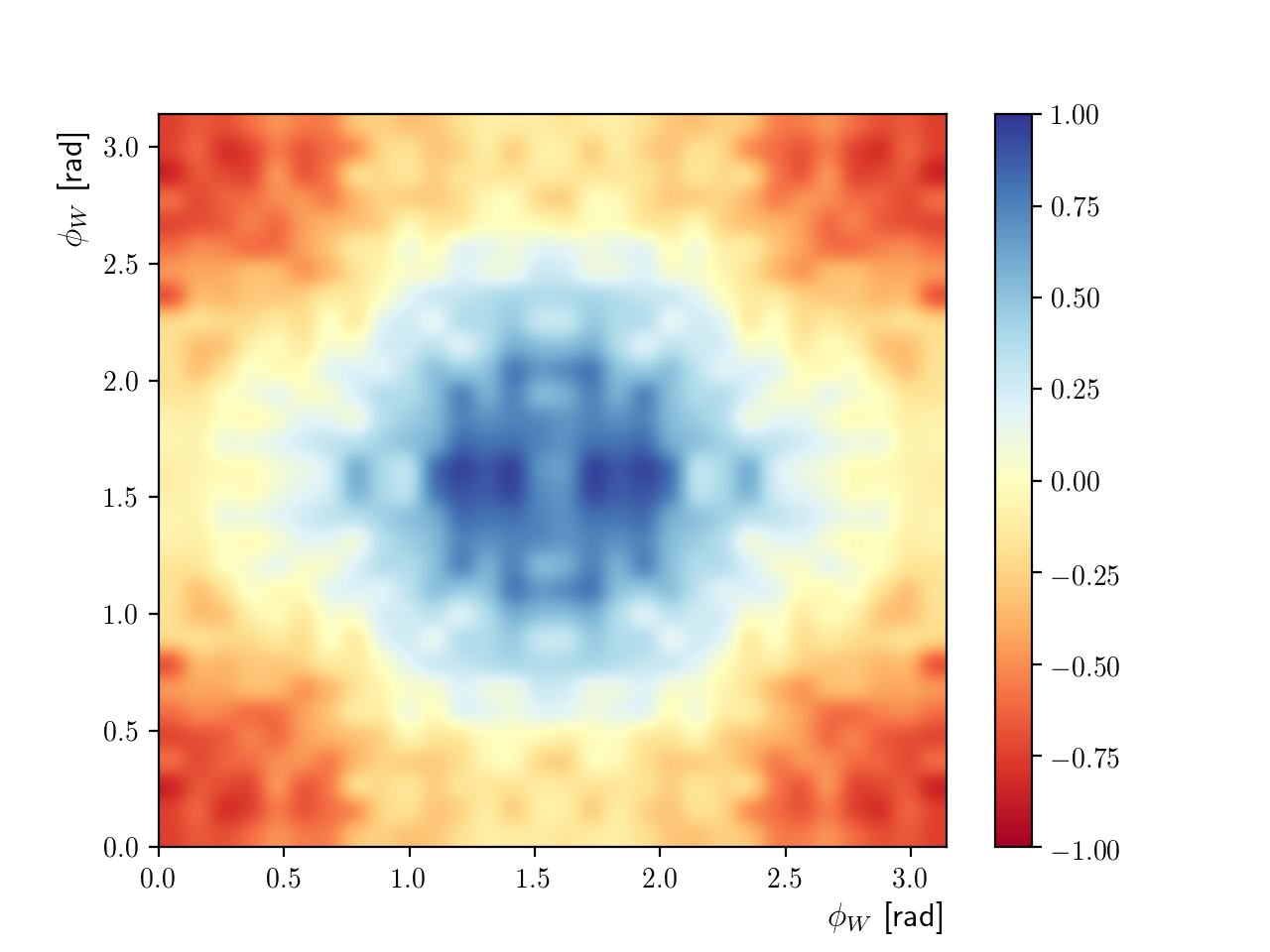} 
		\label{fig:sub1}
	\end{subfigure}
	\begin{subfigure}{.49\textwidth} 
		\vspace{-0.495cm}
		\centering 
		\includegraphics[width=1.1\linewidth, page = 1]{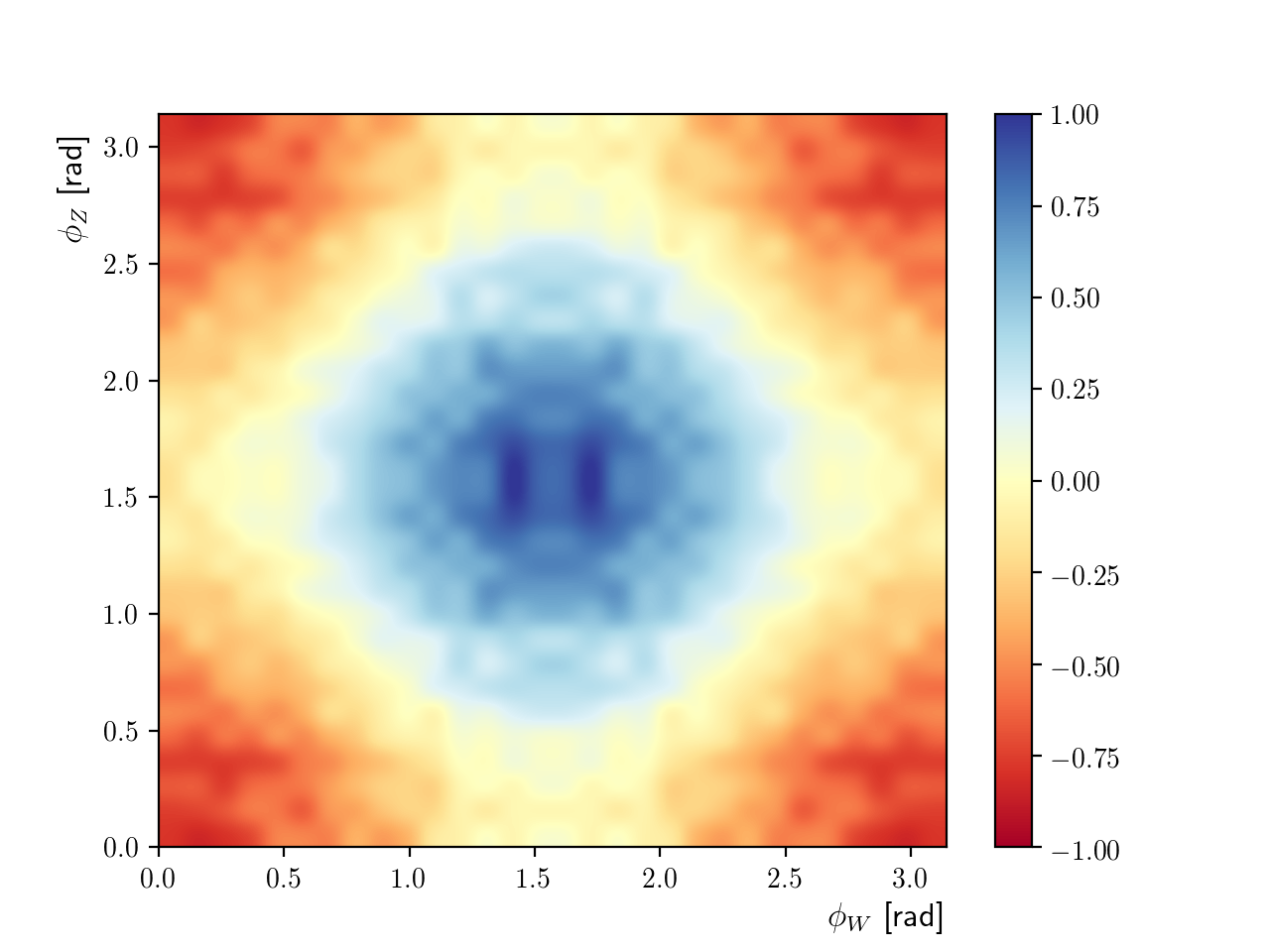} 
	\end{subfigure}
\caption{ Hadronic (left) and Semileptonic (right) SMEFT-SM interference cross-sections after the reconstruction, calorimeter granularization and analysis cuts.
(z axis in arbitrary units). After the foldings, Region A (B) is largely overlapping with the blue (red) region.}
\label{fig:azim_plane} 
\end{figure}

We note that the SM diboson distributions do not exhibit strong dependence on the azimuthal decay
angles (see \cref{appendix:2to4} for parton-level details), and neither do any backgrounds after passing these cuts. Meanwhile, these cuts do not change
the azimuthal behaviour for the BSM interference term. We show both channels for the particular
center-of-mass energy $0.9~\text{TeV} \leq \sqrt{\hat{s}} \leq 1.1$~TeV in Figure~\ref{fig:azim_plane}.

In order to construct an asymmetry variable to maximize the sensitivity to new physics effects, we define two regions of the azimuthal plane:
\beq
\text{Region A}:\, \phi_{V_i} \in \left[\pi/4,\pi/2\right]  \quad
\text{and} \quad
\text{Region B}:\, \phi_{V_i} \in [0,\pi/4]\,,
\eeq 
where $i=1,2$ and $VV$ is $W^+W^-$ in the hadronic case and $W^\pm Z$ in the semileptonic. 
These regions are replicated to the other quadrants after the foldings described in~\cref{sec:substructure}.
Then, the asymmetry is defined as
\beq 
\mathcal{A}(s) \equiv \frac{N_A - N_B}{N_A + N_B}\,,
\eeq 
where $N_{A(B)}$ is the number of events in respective region. 

The number of events in each region is proportional to the total reconstructed cross-section
including SM backgrounds and the single-insertion interference cross-section. While the BSM term has
opposite sign between regions A and B, which add up in the numerator and cancels in denominator, the dominant backgrounds, {\it i.e.} jets ($Z+$jets) for the hadronic 
(semileptonic), are effectively flat in this angular space, resulting in the opposite effect.
Therefore one na\"\i vely expects that the asymmetry should exhibit quadratic growth with $\sqrt{s}$; this is indeed the case when SM diboson production is treated as the sole as background.
However, this is not the largest background for this process. The QCD jets ($Z+$jets) are the dominant background; including them, the asymmetry does grow but it is no longer quadratic. This arises due to the access of the background processes to PDF components with different momentum-fraction dependence compared to those of the diboson processes.

\begin{figure}[!htb]
	\centering 
	\begin{subfigure}{.49\textwidth} 
		\centering
		\includegraphics[width=1.0\linewidth, page = 1]{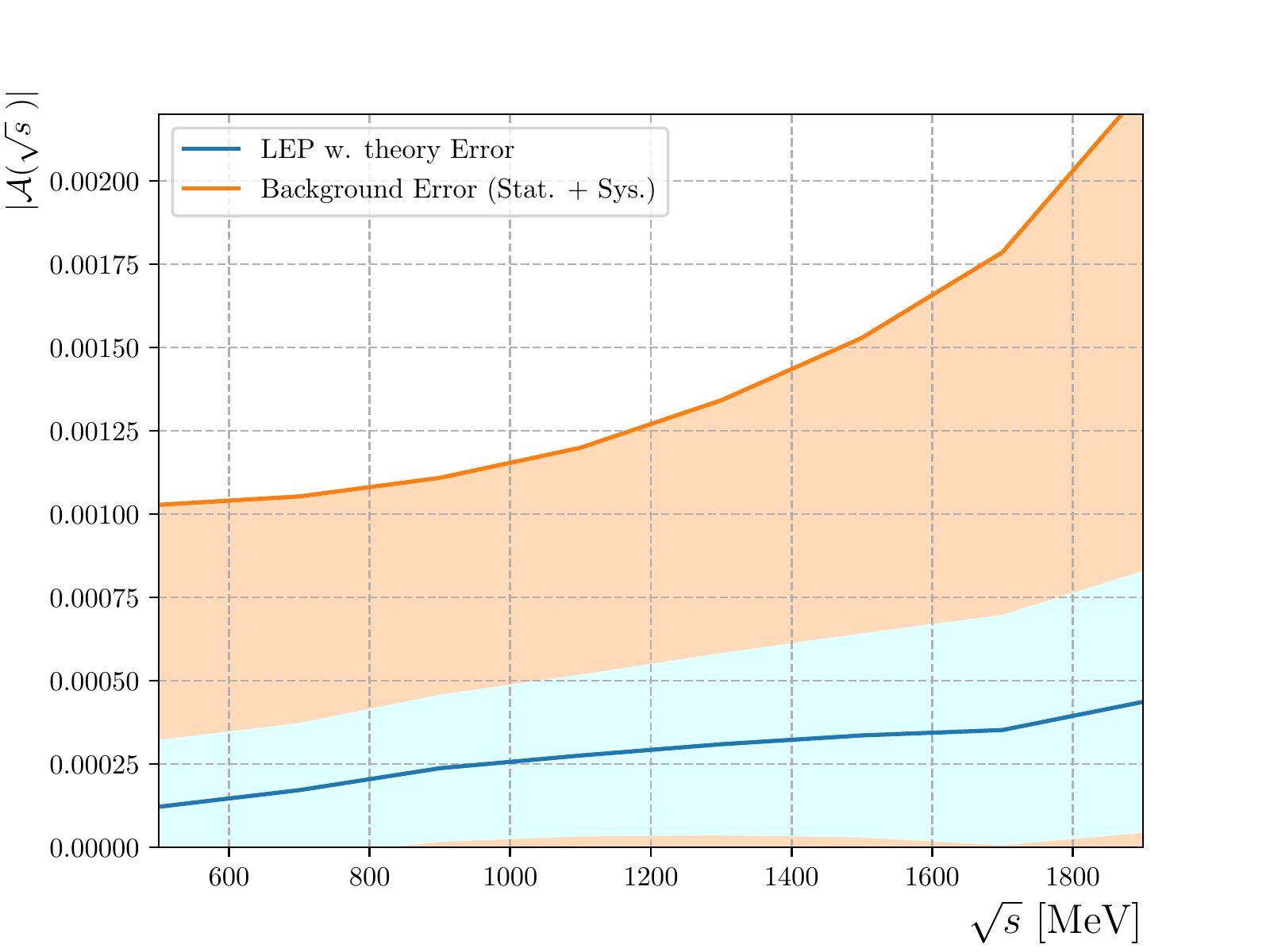} 
		\label{fig:sub1}
	\end{subfigure}
	\begin{subfigure}{.49\textwidth} 
		\centering \includegraphics[width=1.0\linewidth, page = 1]{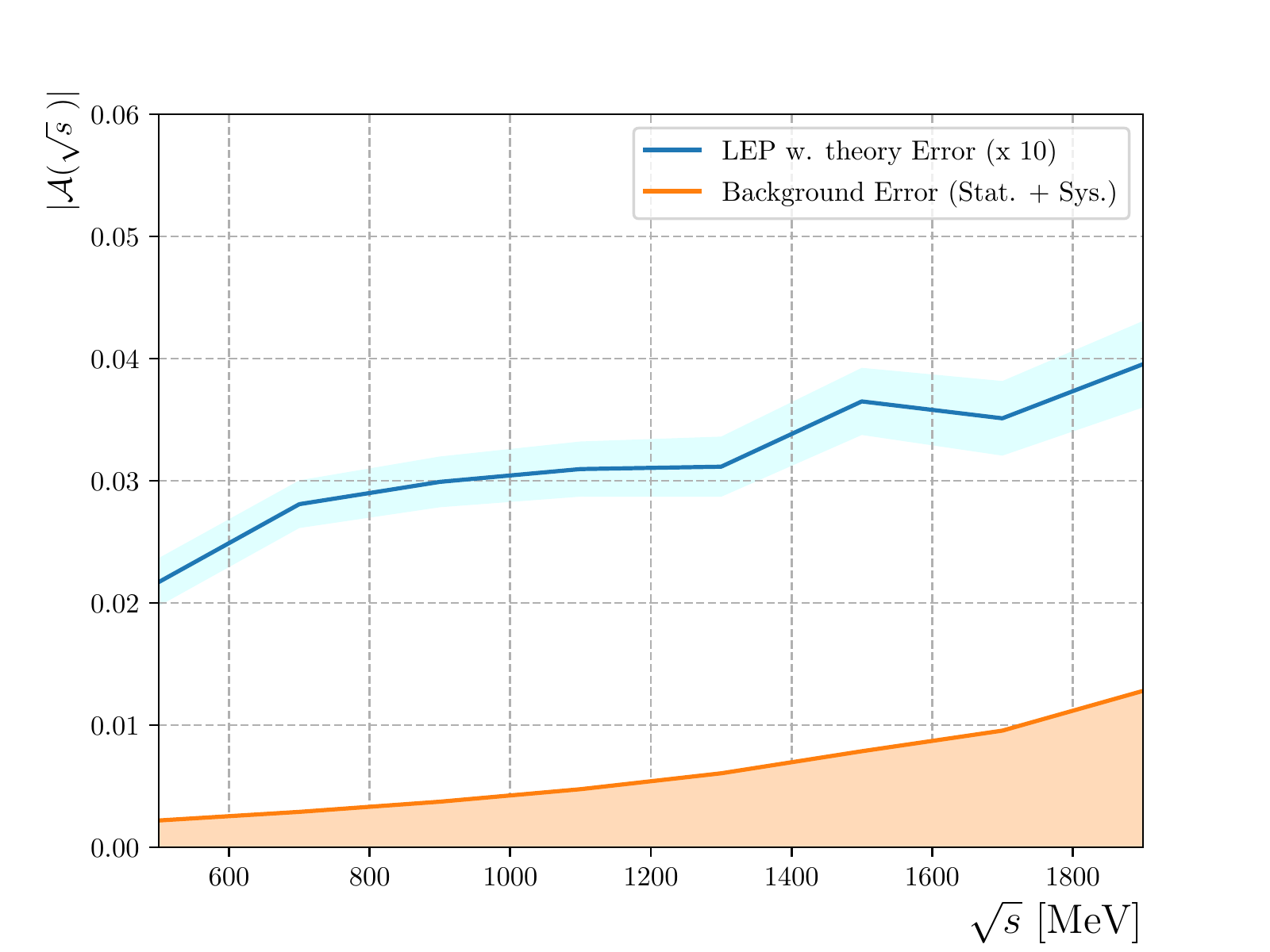} 
		\label{fig:sub2} 
	\end{subfigure}
\caption{Hadronic (left) and Semileptonic (right) absolute asymmetry plots. The blue curves are the asymmetries predicted by the LEP bound value of $C_W$. The statistical 
and systematic (illustratively chosen to be 0.1\% on the asymmetry) uncertainties are shown by the orange shaded region, and the theoretical error due to higher order EFT effects is the blue shaded region around the predicted asymmetry; note that the theoretical error for the semileptonic case has been increased by a factor of 10 for readability.}
\label{fig:asymmetry} 
\end{figure}

In general, the effects of generic SMEFT contributions grow with energy compared to the SM backgrounds, thus it is best to consider the measured asymmetry as a function
of the center-of-mass energy. In Figure~\ref{fig:asymmetry}, we plot the absolute asymmetry, including all backgrounds, for the LEP $2\sigma$ maximal bound\footnote{
The LEPII analysis, much like many others, was driven by the EFT contribution at order $\Lambda^{-4}$, which is incompletely calculated as the square of the EFT contribution and subject to the effects of many other (neglected) operators;
Here, with our analysis cuts, vector bosons are effectively on-shell and the signal prediction is appropriately truncated, yielding an observable which is sensitive to just this operator.} of $\text{max}(|\delta\lambda_Z|) = 0.059$~\cite{Schael:2013ita}, with the illustrative assumed systematic error of $0.1\%$ on the asymmetry and statistical errors assuming an LHC integrated luminosity of $3$ ab$^{-1}$. Note that the theory errors, illustrated as the blue region around the prediction, remain subdominant in all regions of parameter space due to their not contributing directly to the asymmetry, and instead only altering the symmetric background rate.

\section{Detection reach at the LHC} 
\label{sec:results}
Our sensitivity to these asymmetries is driven by the statistical uncertainty of the SM backgrounds;
experimental systematic uncertainties are largely independent of these decay angles, and therefore cancel
in the asymmetry. We thus present results assuming the statistical error is dominant and results
assuming a systematic uncertainty on the asymmetry measurement of 0.1\%, independent of the integrated luminosity.

It is of course necessary to include theoretical errors on the signal model as well. However, we
note that the squared amplitude contribution of various operators do
not themselves contribute to the asymmetry, only to the number of symmetric events in the
background; this holds true for a number of other operator contributions which we have checked explicitly as well. Detailed parton-level calculations are presented in \cref{appendix:2to4,appendix:ops}. Treating the squared SMEFT contributions (with an inflated value of $C_W$ such that $\frac{C_W}{\Lambda^2}=\frac{g}{6m_w^2}$ to remain as conservative as possible) as a theoretical
uncertainty on the signal model, we find that this term produces a structure on the azimuthal plane with peaks around $\phi_V=0$ and $\phi_V=\frac{\pi}{2}$;
these two peaks cancel each other in the asymmetry. This contribution has the same pattern ( $d\sigma= a + b \cos2\phi_{12}\cos2\phi_{34}$ ) as the purely-SM cross section,
but with different coefficients of the two terms. These errors are shown as blue shaded areas around the predicted asymmetries in \cref{fig:asymmetry}.

\begin{figure}[!htb]
\centering
\includegraphics[width=0.85\linewidth, page=1]{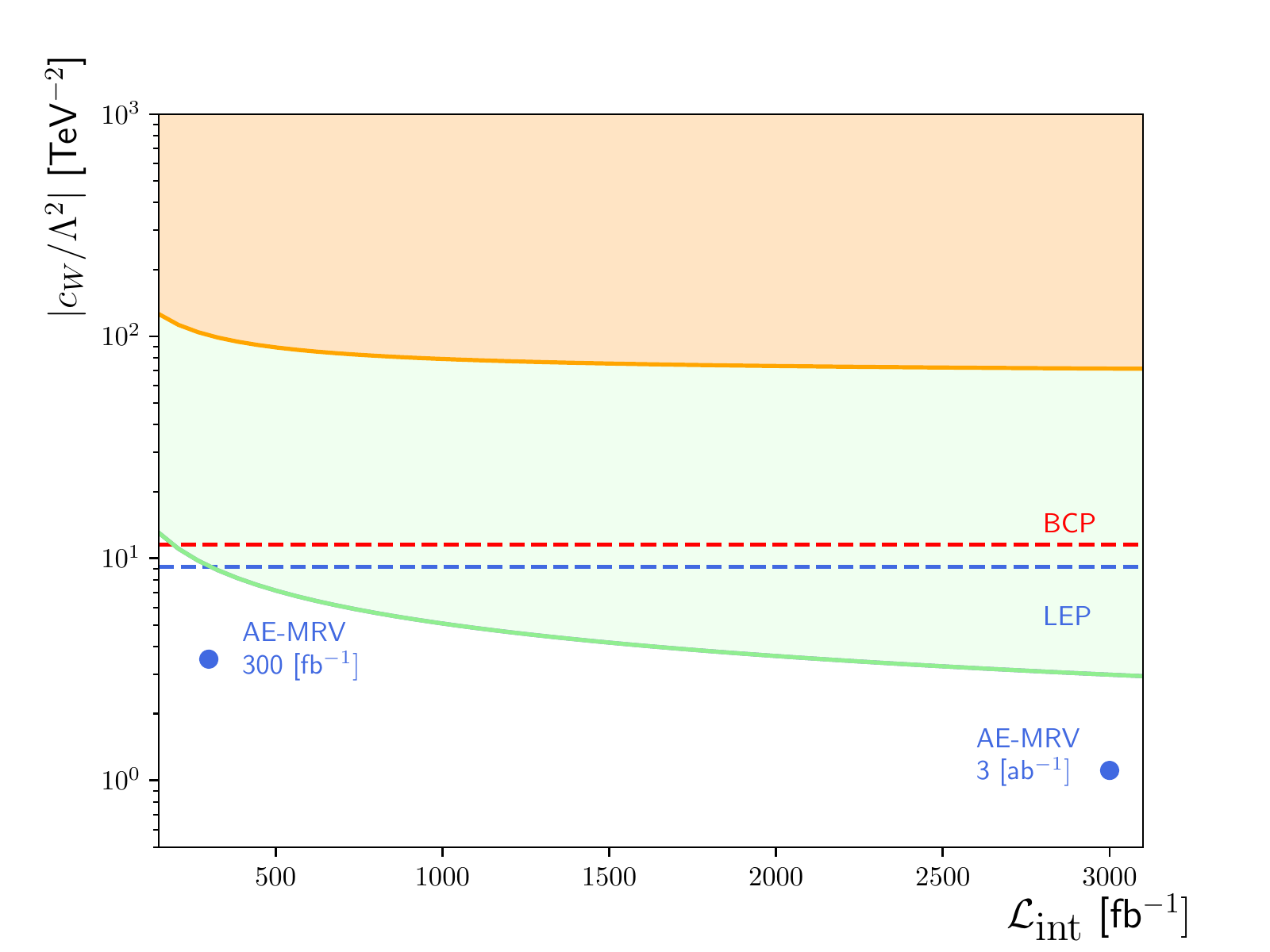} 
\caption{ Hadronic (orange) and Semileptonic (green) detection reach as a function of the integrated luminosity. We also show the LEP $2\sigma$ bound
for this operator, the recent projections (AE-MRV) from Ref.~\cite{Azatov:2017kzw} for $300$fb${}^{-1}$ and $3$ab${}^{-1}$ and the $2\sigma$ bound (BCP) from 
a LHC Run II based global analysis~\cite{Biekotter:2018rhp}, which has a similar treatment of this operator to the LEP analysis.}
\label{fig:reach} 
\end{figure}

Our ultimate reach for detecting these asymmetries at the LHC as a function of integrated luminosity
is shown in Figure~\ref{fig:reach}. The green and orange regions represent the LHC sensitivity to the 
corresponding value of $|c_W/\Lambda^2|$ in the semi-leptonic and fully hadronic channels, respectively. We also indicate the LEP bound, the current LHC $2\sigma$ bound from  Ref.~\cite{Biekotter:2018rhp} (both of which are driven by the theoretically ill-defined $\frac{1}{\Lambda^4}$ EFT contribution), and two prospective reaches, at $300$fb${}^{-1}$ and $3$ab${}^{-1}$, from Ref.~\cite{Azatov:2017kzw} arising from a fully-leptonic final state analysis which neglects the EFT errors due to terms of order $\frac{1}{\Lambda^4}$, but at least does not treat them as signal contributions.
As expected, the more data LHC accumulates, smaller values of $c_W/\Lambda^2$ can be probed.
Even though the total rejection factor for the hadronic channel background is 3 orders of magnitude better than the semileptonic and the signal cross section is larger as well, the QCD multijet background remains problematic. Ultimately, our analysis favors the semileptonic channel in detection reach. We note that, in addition to these two channels, it is also possible to study the azimuthal decay angle of a vector boson produced in association with a leptonically-decaying $W^\pm$ boson using these techniques, but have focused here on final states in which both angles can be reconstructed; single-angle events will not provide a significantly stronger constraint than the two-angle semileptonic study presented here.

\section{Conclusions} 
\label{sec:conc}
We have demonstrated in this article that it is possible to obtain interference measurements in ``non-interfering'' four point
amplitudes using jet substructure techniques. Tagging the subjets with {\it N-subjettiness} and reconstructing their azimuthal decay angles
allows us to probe the anomalous triple gauge coupling generated by the $\mathcal{Q}_{3\text{W}}$ operator. 
With a single-insertion of this operator, patterns exclusive to this operator occur in the azimuthal plane, allowing 
the construction of an asymmetry maximally sensitive to it and that vanishes for other contributions. 

This provides a cross-check of an observable which has been consistently calculated in the SMEFT, including
honest theory error appraisals. Thus, it continues the work toward developing the needed machinery
to perform a fully-general, model-independent analysis of the implications of precision SM
measurements at the LHC and elsewhere on heavy new physics. It is interesting to note that, in the at least semi-hadronic signals considered here, the theory errors remain always subdominant to the SM statistical errors, which stands in stark contrast to the other signatures studied consistently in the SMEFT. This arises because of the unique nature of these azimuthal interference patters even within the SMEFT itself, preventing other operators from contributing to the constructed asymmetry. The errors in a purely leptonic analysis which properly truncates its signal at leading order in SMEFT effects are similarly suppressed.

It is particularly interesting that these techniques give access to observables which measure exactly
one term in the Warsaw basis of the SMEFT Lagrangian, since other observables require a higgs-vev insertion and,
since they do not have an extra derivative, they do not flip one of the diboson helicities. This lack of helicity flip
makes an azimuthal distribution which is not doubly-odd, therefore cancels in the asymmetry at leading order.

Another possibility of this technique is to construct a similar search for
these non-interference effects in the QCD sector, as initially proposed in \cite{Dixon:1993xd} and recently revisited in \cite{Hirschi:2018etq}. We shall explore whether these substructure techniques now make this azimuthal angular search possible in a future publication.

\section*{Acknowledgments} We thank Camila Machado, Lucia Massetti and Peter Berta for enlightening discussions.
The work of RA, and partly that of WS, was supported by the Alexander von Humboldt Foundation, in the framework of the Sofja
Kovalevskaja Award 2016, endowed by the German Federal Ministry of Education and Research.

\bibliography{Sources}
\bibliographystyle{JHEP}
\newpage

\appendix \section{Two-to-four partonic cross section calculations} 
\label{appendix:2to4} 
The SM diboson production matrix elements at leading order in the squared center of mass energy 
$s$ are given by (for $u\bar{u} \rar W^+W^-$, other cases are similar)~\cite{Falkowski:2016cxu}

\begin{align*}
\M^{\text{SM}}(-+ \rar 00) &= \frac{3 g^2_L + g^2_Y}{12} \sin \theta + \OO(m^2_W /s), & \M^{\text{SM}}(+- \rar 00) &= - \frac{g^2_Y}{3} \sin \theta + \OO(m^2_W /s),\\
\M^{\text{SM}}(-+ \rar \pm \mp) &= - \frac{\mp 1 + \cos \theta}{1+\cos \theta} \frac{g^2_L}{2} \sin \theta, & \M^{\text{SM}}(+- \rar \pm \mp) &= 0,
\end{align*}
for the BSM, and for the single-insertion of the effective coupling $\delta\lambda_Z$:

\begin{align*}
\M^{\text{BSM}}(-+ \rar \pm 0) &= \frac{\sqrt{s}}{M_W} \frac{g_L}{12\sqrt{2}} (\pm 1 - \cos \theta) 3 \delta\lambda_Z, & \M^{\text{BSM}}(-+ \rar \pm \pm) &= \frac{s}{M^2_W} \frac{g^2_L}{4} \sin\theta \delta\lambda_Z.
\end{align*}

The vector boson decay amplitudes for helicity $\lambda$ can be decomposed as:
\beq
	\A^{\lambda} = (g_{\psi}\sqrt{2s}/s_w) e^{\lambda \phi_{ij}} d_\lambda(\theta_{ij})
\eeq
where $d_\lambda(\theta_{ij})$ are the Wigner functions $d_\pm(\theta_{ij}) = (1 \mp \cos\theta_{ij})/2$ and $d_0(\theta_{ij}) = (\sin\theta_{ij})/\sqrt{2}$.

\begin{table}[!htb] 
\centering 
\begin{tabular}{|c|c|}  \hline
$\lambda$ &$ \A_{\lambda}/(g_{\psi}\sqrt{2
s}/s_w)$ \\ \hline \hline $0$ &$ -\sqrt{2} \sin \theta_{ij} $\\ $+$ &$ (1-\cos \theta_{ij})
e^{i\phi_{ij}} $\\ $-$ &$ (1+\cos \theta_{ij}) e^{-i\phi_{ij}} $\\\hline \end{tabular} \end{table}

With the above constituents, we can define the total partonic spin-averaged two-to-four cross
section
\beq \sigma (s) = \int \frac{\sum |\M|^2}{8s} \frac{ds_{12} ds_{34}}{(2\pi)^2} \left[ \frac{1}{8\pi}
\frac{d\cos\theta_{12}}{2}\frac{d\phi_{12}}{2\pi} \right] \left[ \frac{1}{8\pi}
\frac{d\cos\theta_{34}}{2}\frac{d\phi_{34}}{2\pi} \right] \left[ \frac{\beta}{8\pi}
\frac{d\cos\theta}{2}\frac{d\phi}{2\pi} \right], \eeq
with
\bea \sum |\M|^2 &=& |D_W(s_{12}) D_W(s_{34})|^2 \sum_{\lambda_{12}\lambda'_{12}}
\sum_{\lambda_{34}\lambda'_{34}} \lp \A^{\lambda_{12}} \A^{\lambda_{34}} \rp \lp \A^{\lambda'_{12}}
\A^{\lambda'_{34}} \rp^* \\\nonumber &~& \sum_{\lambda_{+}\lambda_{-}} \lp \M_{q\bar{q} \rar W
W}^{\lambda_+\lambda_-,\lambda_{12}\lambda_{34}}\rp \lp \M_{q\bar{q} \rar W
W}^{\lambda_+\lambda_-,\lambda'_{12}\lambda'_{34}}\rp^*, \eea
where $D_W(s)$ is the $W$ boson propagator stripped of Lorentz structure, $\lambda_\pm$ are initial
state quark polarizations, and $\lambda^{(\prime)}_{12,34}$ are the polarizations of the
intermediate, decaying $W$ bosons. Crucially, because these particles are internal to the process,
the summation over their helicities takes place independently in the amplitude and its complex
conjugate.

\subsection{Single-insertion interference}

The total BSM amplitude including decays is 
\beq \M^{\text{BSM}} (-+ \rar \pm\pm) \sim 
\lp -\frac{\delta\lambda_Z s}{m_W^2} \rp
\lp   \frac{g^2_L}{2} \sin\theta \rp (1\mp \cos\theta_{12})(1\mp \cos\theta_{34}) e^{\pm
i(\phi_{12} -\phi_{34})}, 
\eeq
and the interference with the SM amplitude can be written as 
\beq \sum |\M|^2 \sim \sum_{\lambda_{12}\lambda_{34}} 2
\text{Re}\[ \M^{\text{SM}}_{-+,\lambda_{12}\lambda_{34}} (\M^{\text{BSM}}_{-+,--})^* +
\M^{\text{SM}}_{-+,\lambda_{12}\lambda_{34}} (\M^{\text{BSM}}_{-+,++})^* \]. \eeq

In the equation above, two BSM three SM helicities configuration contribute for the sum. The azimuthal behavior of SM 
and BSM can be seen in table below, where $\Delta\phi=\phi_{12}-\phi_{34}$. 
\begin{table}[!ht] 
	\label{tab:azimuthal_single}
	\centering 
	\begin{tabular}{|cc|c|c|} \hline
	$(\lambda'_{12}\lambda'_{34})$ & $(\lambda_{12}\lambda_{34})$ & polar & azimuthal \\\hline \hline 
	$(++)$ & $(00)$  & $(1-\cos\theta_{12})(1-\cos\theta_{34})(2 \sin\theta_{12}\sin\theta_{34})$ &  $2 \cos \Delta\phi$ \\ 
	       & $(+-)$  & $(1-\cos\theta_{12})^2 \sin^2\theta_{34}$                                &  $2 \cos 2\phi_{34}$ \\
	       & $(-+)$  & $(1-\cos\theta_{34})^2 \sin^2\theta_{12}$                        &  $2 \cos 2\phi_{12}$ \\ \hline
	$(--)$ & $(00)$  & $(1+\cos\theta_{12})(1+\cos\theta_{34})(2 \sin\theta_{12}\sin\theta_{34})$ &  $2 \cos \Delta\phi$ \\ 
	       & $(+-)$  & $(1+\cos\theta_{34})^2 \sin^2\theta_{12}$                                &  $2 \cos 2\phi_{34}$ \\
	       & $(-+)$  & $(1+\cos\theta_{12})^2 \sin^2\theta_{34}$                                  &  $2 \cos 2\phi_{12}$ \\ \hline
	\end{tabular}
\end{table} 

One should notice here that the coefficient of $\cos2\phi_{12} (2\phi_{34})$ goes to zero whenever $\theta_{34} (\theta_{12})
\rar 0,\pi$ and vice versa. This means that the interference vanishes again whenever
$\theta_{(12),(34)} \rar 0,\pi$, i.e. in the limit where decay products of either $W$ boson are
either collinear or anti-collinear with the $W$ momentum. This is intuitive to understand as arising
from the fact that in this limit the decay angle, which determines whether the interference is
constructive or destructive in the normal case, cannot be defined leading to an effectively-longitudinal polar angle distribution of the vector boson decay products, even though it is the result of interference of two distinct transverse polarizations.

Integrated over the polar angles of $W$ pair production and $W$ boson decays, the interference term
becomes:

\bea 
\frac{d\sigma_{\text{int}}(s)}{d\phi_{12}d\phi_{34}} \sim
\lp-\frac{\delta\lambda_Z s}{m_W^2}\rp 
\left\{ [g_L^2 ( 3g_L^3 + g_Y^2) \frac{\pi}{2} ] \cos{\Delta\phi} + g_L^4 ( \cos 2\phi_{12} + \cos 2\phi_{34}) \right\} 
\eea
where the first term vanish after the proper foldings.

\subsection{SM self-interference} 
In the purely SM case, the squared amplitude is given by
\bea 
\sum |\M|^2 &=&
\sum_{\lambda_{12}\lambda'_{12}}\sum_{\lambda_{34}\lambda'_{34} }\sum_{\lambda_+\lambda_-}
\M^{\text{SM}}_{\lambda_+\lambda_-,\lambda_{12}\lambda_{34}}
(\M^{\text{SM}}_{\lambda_+\lambda_-,\lambda_{12}\lambda_{34}})^* \\\nonumber
 &=&\sum_{\lambda_{12}\lambda'_{12}}\sum_{\lambda_{34}\lambda'_{34} }
\M^{\text{SM}}_{+-,\lambda_{12}\lambda_{34}} (\M^{\text{SM}}_{+-,\lambda_{12}\lambda_{34}})^* +
\M^{\text{SM}}_{-+,\lambda_{12}\lambda_{34}} (\M^{\text{SM}}_{-+,\lambda_{12}\lambda_{34}})^* 
\eea
The first term, coupling to right-handed fermions, produces only longitudinal gauge bosons at
leading order. Thus, it has no azimuthal dependence, as there is no nontrivial combination of
helicities to interfere. The helicity interference structure of the second term is:

\begin{table}[!ht] 
	\centering 
	\begin{tabular}{|cc|c|c|} \hline 
	$(\lambda'_{12}\lambda'_{34})$ & $(\lambda_{12}\lambda_{34})$ & polar & azimuthal\\\hline \hline 
	$(00)$ & $(00)$ & $4(\sin\theta_{12})^2 (\sin\theta_{34})^2$ & 1  \\ 
	       & $(+-)$ & $2(\sin\theta_{12}) (\sin\theta_{34}) (1 - \cos\theta_{12})(1-\cos\theta_{34})$ & $e^{-i\Delta \phi}$  \\
	       & $(-+)$ & $2(\sin\theta_{12}) (\sin\theta_{34}) (1 + \cos\theta_{12})(1+\cos\theta_{34})$ & $e^{+i\Delta \phi}$ \\ \hline
	$(+-)$ & $(00)$ & $2(\sin\theta_{12}) (\sin\theta_{34}) (1 - \cos\theta_{12})(1-\cos\theta_{34})$ & $e^{+i\Delta \phi}$  \\ 
	       & $(+-)$ & $(1-\cos\theta_{12})^2 (1+\cos\theta_{34})^2$ & 1 \\
	       & $(-+)$ & $(1-\cos\theta^2_{12}) (1-\cos^2\theta_{34})$ & $e^{+2i\Delta\phi}$   \\ \hline
	$(-+)$ & $(00)$ & $2(\sin\theta_{12}) (\sin\theta_{34}) (1 + \cos\theta_{12})(1+\cos\theta_{34})$ & $e^{-i\Delta \phi}$  \\ 
	       & $(+-)$ & $(1-\cos\theta^2_{12}) (1-\cos^2\theta_{34})$ & $e^{-2i\Delta\phi}$ \\ 
	       & $(-+)$ & $(1+\cos\theta_{12})^2 (1-\cos\theta_{34})^2$ & 1 \\ \hline 
	\end{tabular} 
\end{table}

Note that in this case there are combinations of helicities which allow for nonvanishing forward
decay amplitudes, in contrast to the interference case. This means that there is a higher
probability of a $W$ boson decaying into something which appears to be a simple QCD jet, yielding
different tagging efficiencies for SM diboson pair production as compared to single-insertion of
$\delta\lambda_Z$ diboson production. This is again due to the SM amplitude having a large contribution from straightforward transverse polarizations, something forbidden by the non-interference effects for the single-insertion SMEFT case.

Summing and integrating over the polar decay angles $\theta_{12}$ and $\theta_{34}$,

\bea 
\sum|\M|^2 &\sim& \frac{64}{9}\left[ \lp \frac{g^2_Y \sin
\theta}{3}\rp^2 + \lp\frac{3 g^2_L + g^2_Y}{12} \sin \theta \rp^2 \right] \nonumber +
\frac{\pi^2}{2} \left[ \lp \frac{1-\cos\theta}{1+\cos\theta}\frac{g^2_L}{2}\sin\theta \rp^2 + \lp
\frac{g^2_L}{2}\sin\theta\rp^2 \right] \\\nonumber &+& \frac{512}{9} \lp
\frac{1-\cos\theta}{1+\cos\theta}\frac{g^2_L}{2}\sin\theta \rp \lp \frac{g^2_L}{2}\sin\theta\rp
\cos2\phi_{12}\cos2\phi_{34}. 
\eea
Na\"\i vely, one expects a divergence whenever $\cos\theta = - 1$ which corresponds to the coulomb pole for $t$-channel quark exchange, 
but integrating over the detector acceptance $|\eta| \leq 4.9$ regulates this divergence. 
The constant term dominates over the term with nontrivial azimuthal structure. 
Therefore, for practical terms one can consider the SM self-interference to be effectively independent of azimuthal decay angles.

\subsection{EFT self-interference} 
Finally, for the $\delta\lambda_Z^2$ cross section, we have
\bea \sum |\M|^2
&=& \sum_{\lambda_{12}\lambda'_{12}}\sum_{\lambda_{34}\lambda'_{34} }\sum_{\lambda_+\lambda_-}
\M^{\text{BSM}}_{\lambda_+\lambda_-,\lambda_{12}\lambda_{34}}
(\M^{\text{BSM}}_{\lambda_+\lambda_-,\lambda_{12}\lambda_{34}})^* \\\nonumber &=& \lp
\frac{s}{m_W^2} \frac{g^2_L}{4} \sin\theta \delta\lambda_Z \rp^2
\sum_{\lambda_{12}\lambda'_{12}}\sum_{\lambda_{34}\lambda'_{34} }
\M_{\lambda_{12}}\M_{\lambda_{34}}(\M_{\lambda'_{12}}\M_{\lambda'_{34}})^* 
\eea
The structure of the individual terms of this sum with respect to decay angles is:
\begin{table}[!ht] 
	\centering 
	\begin{tabular}{|cc|cc|} \hline  
	$(\lambda'_{12}\lambda'_{34})$ & $(\lambda_{12}\lambda_{34})$ & polar &  azimuthal \\\hline \hline
	$(++)$ & $(++)$ & $(1-\cos\theta_{12})^2 (1-\cos\theta_{34})^2$ & no \\ 
	       & $(--)$ & $\sin^2\theta_{12} \sin^2\theta_{34}$ & $e^{2i\Delta\phi}$\\ 
	$(--)$ & $(++)$ & $\sin^2\theta_{12} \sin^2\theta_{34}$ & $e^{-2i\Delta \phi}$\\ 
	       & $(--)$ & $(1+\cos\theta_{12})^2 (1+\cos\theta_{34})^2$ & no \\ \hline 
	\end{tabular} 
\end{table}

Note again that there exist here terms which allow for forward $W$ decays. Integrating over polar
decay angles leads to
\beq 
\sum|\M^{\text{BSM}}|^2 \sim \lp \frac{s}{m^2_W} g_L^2 \delta\lambda_Z \rp^2 \lp
\frac{32}{27} \rp \lp 1 + 4 \cos2\phi_{12}\cos2\phi_{34}\rp 
\eeq

We can see that it is composed of the same azimuthal structures as the pure SM case, but with differing weights
between the constant and $\phi_{ij}$-dependent terms.

\section{Effects of additional operators}
\label{appendix:ops}

Considering the effects of the other contributions to
triple gauge boson couplings $\delta g_{1,Z}$, $\delta \kappa_Z$ and $\delta \kappa_{\gamma}$, we
have the following LO amplitudes: 
\bea 
	\M^{\text{BSM}}(-+ \rar 00) &=& \frac{\sqrt{s}}{m_W} \frac{g_L}{12} \sin\theta [ -3\delta\kappa_Z - 4s^2_W ( \delta \kappa_{\gamma} - \delta\kappa_Z)]\\\nonumber 
	\M^{\text{BSM}}(+- \rar 00) &=& \frac{s}{m^2_W} \frac{g^2_L}{4} \sin\theta [2s^2_W ( \delta \kappa_{\gamma} - \delta\kappa_Z)],
\eea 
where we have replaced
$\delta\kappa_\gamma$ using the relation $\delta \kappa_Z = \delta g_{1,Z} - s^2_W
\delta \kappa_{\gamma}$, which holds for our chosen SMEFT input scheme. Interfering with the SM amplitudes with $\lambda_{12},\lambda_{34} =
\pm\mp$ will give terms as $\cos{\left(\phi_{12} + \phi_{34}\right)}$ which cancel after the two foldings.
One should notice that also exist the possibility of these operators generating the same transverse
amplitudes as the SM: $\left(\lambda_{12},\lambda_{34} = \pm\mp \right)$, however this comes at $\mathcal{O}(s^0)$ and
no energy growth relative to the SM cross section is expected.
The terms $\lambda_{12},\lambda_{34} = 00$ are of course independent of the azimuthal decay angles. The
total squared amplitude integrated over polar decay angles is 
\bea 
\sum|\M|^2 &\sim& \lp \frac{s}{m^2_W} g_L^2 \sin\theta \rp^2 \lp \frac{512}{324}\rp \\\nonumber 
			&~& \left[ -\delta g_{1,Z} \lp \frac{9g^2_L + g^2_Y}{3} \rp + \delta\kappa_Z \lp (4s_W + 1) \frac{3g^2_L + g^2_Y}{12} - 2 g^2_L c^2_W \rp\right].
\eea 

An additional potential source of azimuthal behavior that could fake this signal is singly-resonant diagrams involving a four-fermion operator. The azimuthal behavior of the non-resonant fermion pair in this case would not be dictated by the helicity of an on-shell vector boson, and as such is more challenging to calculate analytically. We have numerically explored these processes and found no operator which contributes to the asymmetries studied here.

\section{Signal and Background Events Tables}
In this Appendix we present some illustrative numbers for ($N_A$,\,$N_B$), as defined in Section~\ref{sec:search}, as a function
of the Wilson coefficients for linear and quadratic in Table~\ref{tab:NAB_Cw}. Moreover, we present the same numbers for various backgrounds in Table~\ref{tab:NAB_bkg}. All the numbers are for an integrated Luminosity of $300\,$[fb]$^{-1}$.

\begin{table}[!ht] 
\centering 
	\scalebox{1.0}{
	\begin{tabular}{|c|c||c|} \hline
	   &Hadronic & Semileptonic\\ \hline\hline
	  linear   &    (2.22, -1.95)$\,10^{3} \times \delta\lambda_Z$    & (1.28, -1.28)$\, 10^{3} \times \delta\lambda_Z$ \\ \hline
	  quadratic &(3.45,3.53)$\,10^{6} \times (\delta\lambda_Z)^2$  &  (0.68,0.67)$\, 10^{6} \times (\delta\lambda_Z)^2$ \\ \hline
	\end{tabular}}
	\caption{The number of events $(N_A\,,N_B)$ for linear and quadratic of the cross-section.
	One could easily obtain the same numbers for different Wilson Coefficient values just rescaling the showed numbers with the aid of Eq.~\ref{eq:deltaCs} in Section~\ref{subsec:TGC}. The numbers of $N_B$ for the linear part can be negative since the interference cross-section has oppostive sign in these two regions. Note that the numbers of events from the quadratic contributions are equal up to statistical errors in Monte Carlo samples, justifying our claim that theoretical errrors have reduced impact on this asymmetry observable.}
	\label{tab:NAB_Cw}
\end{table}

\begin{table}[!ht] 
\centering 
	\scalebox{0.87}{
	\begin{tabular}{|c|c|c||c|c|c|} \hline
	  \multicolumn{3}{|c||}{Hadronic}                    &\multicolumn{3}{c}{Semileptonic} \vline\\ \hline\hline
	   SM &  jets         & $t\bar{t}$                                    &  SM & Z+jets          & $t\bar{t}$            \\ \hline
	  (2.95,2.97)$\,10^{3}$  & (4.60,4.49)$\,10^{4}$ &(4.28,4.32)$\,10^{5}$    & (1.26,1.26)$\,10^{3}$  & (2.88,2.88)$\, 10^{3}$ &  (0.46,0.46)$\, 10^{3}$ \\ \hline
	\end{tabular}}
	\caption{Number of events for some of the hadronic and semileptonic backgrounds $(N_A\,,N_B)$, after histogram normalisations and cuts.As expected, both regions A and B have equal numbers of events (within statistical errors of our Monte Carlo sampling) and the same sign, making their expected contribution to the defined asymmetry vanish.}
	\label{tab:NAB_bkg}
\end{table}

\end{document}